\documentstyle[prl,twocolumn,aps]{revtex}
\input{epsf}
\begin{document}
\draft
\twocolumn[\hsize\textwidth\columnwidth\hsize\csname @twocolumnfalse\endcsname
\title{Quantum computing of quantum chaos and imperfection effects}

\author{Pil Hun Song$^{(a)}$ and Dima L. Shepelyansky$^{(b)}$}

\address {$^{(a)}$Max-Planck-Institut f\"{u}r Physik komplexer
Systeme, N\"{o}thnitzer Str.~38, D-01187 Dresden, Germany\\
$^{(b)}$Laboratoire de Physique Quantique, UMR 5626 du CNRS, 
Universit\'e Paul Sabatier, F-31062 Toulouse Cedex 4, France}

%\date{\today}
\date{September 1, 2000}

\maketitle

\begin{abstract}
We study numerically the imperfection effects in the
quantum computing of the kicked rotator model
in the regime of quantum chaos.  It is shown
that there are two types of physical characteristics:
for one of them the quantum computation errors grow 
exponentially with the number of qubits in the computer 
while for the other the growth is polynomial.  
Certain similarity between classical and quantum
computing errors is also discussed.
\end{abstract}
\vskip1pc]

%\begin{multicols}{2}
\narrowtext

A great interest to quantum computers has been generated 
recently by prominent theoretical results  
and impressive experimental progress which allowed to realise
operations with a few qubits (see \cite{steane} for a review).
The most striking theoretical advantage is an enormous parallelism of 
quantum computing.  Using the Shor algorithm \cite{shor} the 
factorization of large numbers can be done
exponentially faster on a quantum computer
than by any known algorithm on a classical computer.
Also a search of an item in a long list is much faster on a
quantum computer as shown by Grover \cite{grover}.
Experimentally a variety of physical systems is considered
for realisation of one qubit, viewed as a two level system,
and controlled coupling between a few qubits that forms
the basis for realisation of a quantum computer.
These systems include ion traps \cite{zoller,sor2}, 
nuclear magnetic resonance systems
\cite{nmr}, nuclear spins with interaction controlled electronically
\cite{vagner,kane} or by laser pulses \cite{bowden}, electrons 
in quantum dots \cite{loss}, Cooper pair boxes \cite{cooper}, 
optical lattices \cite{lattice} and electrons floating on liquid 
helium \cite{helium}.  As a result, a two-qubit gate has been experimentally
realized with cold ions \cite{monroe}, and the Grover algorithm
has been performed for three qubits made from nuclear spins in a molecule 
\cite{3q}. 

It is clear that in any realistic realisation of a quantum computer
a special attention should be paid to the imperfection effects.
Indeed, the imperfections are always present and they
in principle may seriously modify the computation results 
comparing to the algorithms based on ideal qubit operations.
At present the imperfection effects are tested in the numerical 
simulations of the quantum Fourier transform (QFT) \cite{zoller}
and the Shor algorithm factorization of 15 \cite{paz,zurek}.
The obtained results look to be promising for
the quantum computing indicating that a small amount
of noise does not change strongly the computations \cite{zoller}
even if in some cases only rather low level of noise
is tolerable \cite{zurek}. However, for different 
reasons these studies do not allow to obtain
analytical estimates of a tolerable imperfection level
for a large number of qubits $n$. Indeed, the Shor
algorithm is rather complicated and the capability of
nowadays computers become too restrictive \cite{paz,zurek}.
Recently the effects of static imperfections
on the the stability of quantum computer hardware 
have been determined for a broad regime of parameters
and it has been shown that the quantum hardware 
is sufficiently robust \cite{gs}. However, these results
cannot be directly generalised for a specific quantum algorithm
operating in time.
 
In the view of importance of imperfection effects 
we analyse in this paper their influence
on a quantum computation of quantum chaos evolution in time.
The quantum chaos in time-dependent systems was studied
intensively during last two decades 
and it has been understood that the quantum interference
can lead to dynamical localisation of classical diffusive
excitation in a close analogy with the Anderson
localisation in a random potential \cite{fishman,chirikov89,izrailev}.
The study of such systems should represent a serious test
for quantum computing. Indeed, in a classically chaotic
system the numerical errors grow exponentially with time
due to exponential local instability of motion
which leads to chaotic diffusion in the phase space. 
In the quantum  case  the error growth
is not so strong  \cite{ds83} but still the dynamical localisation
of quantum chaos remains rather sensitive to
external perturbations and noise \cite{ds83,ott}.
In the view of quantum computing, our work is relevant to 
the situation when a problem is engaged with iterative quantum gate 
operations on a quantum computer,
e.g. the Grover's algorithm\cite{grover}.  

To investigate the imperfection effects on quantum computing
we choose the kicked rotator model introduced in
\cite{casati79}. This model represents the main features
of time-dependent quantum chaos and had been studied extensively
in numerical simulations \cite{chirikov89,izrailev}
and experiments with cold atoms \cite{raizen,amman}.
The unitary evolution operator $\hat U$ over the period $T$ of 
the perturbation is given by
\begin{eqnarray} 
\label{qmap}
\bar{\psi} = \hat{U} \psi =  e^{-ik\cos{\hat{\theta}}}
 e^{-iT\hat{n}^2/2} \psi,
\end{eqnarray}
where $\hbar=1$ so that
the commutator is $[\hat{n},\hat{\theta}]=-i$ and the
classical limit corresponds to $k \rightarrow \infty$, 
$T \rightarrow 0$ while the classical chaos parameter $K=kT$ 
remains constant. The operator $\hat{U}$ is given
by the product of two unitary operators representing  kick 
$\hat{U}_k=\exp(-i k\cos{\hat{\theta}})$
and free rotation $\hat{U}_T=\exp(-iT\hat{n}^2/2)$.
The dynamics is considered on $N$ quantum levels
with periodic boundary conditions.
In the classical limit the dynamics is described by the Chirikov standard map:
\begin{eqnarray} 
\label{cmap}
\bar{n} = n + k \sin{  \theta }; \;\;\; 
\bar{\theta} = \theta + T \bar{n}.
\end{eqnarray}
For $K > 0.9716$ the global chaos sets in with the  diffusive growth
$n^2 = D t$ where $t$ is given in number of kicks and 
the diffusion rate is $D \approx k^2/2$ for $K > 4.5$ \cite{izrailev}.
The quantum interference leads to suppression of this diffusion
after a time scale $t^* \approx D$ and exponential localisation
of the eigenstates of $\hat{U}$ operator  with the localisation length
$l \approx D/2$ \cite{chirikov89,izrailev}.
\vskip -0.3cm
\begin{figure}
\epsfxsize=3.2in
\epsfysize=2.6in
\epsffile{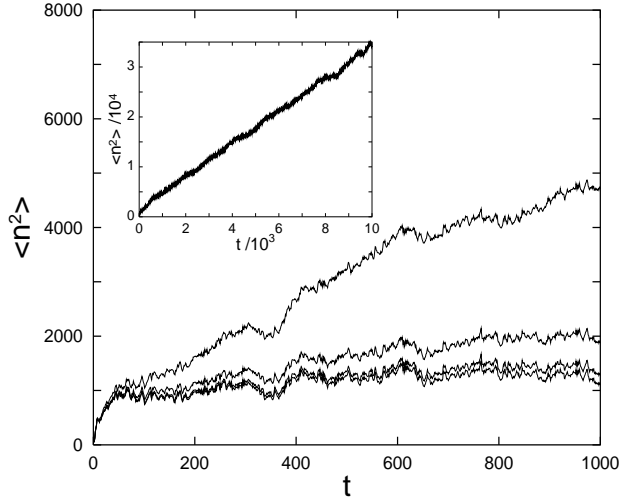}
%\vglue -0.3cm
%\medskip
\caption{Dependence of the second moment $\langle n^2\rangle$ on time $t$
for different imperfection  strength $\epsilon$ in quantum computing
and different number of qubits $n_q$. Curves are for $n_q=13, 12, 11$ 
at $\epsilon =10^{-4}$ from top to bottom and the lowest curve
is for $\epsilon=0$ being the same for $n_q=13, 12, 11$.
Here $k=10, K=5$ and  at $t=0$ all probability is at $n=0$. 
Inset shows the upper curve up to larger times. 
}
\label{fig1}
\end{figure}

The most efficient way of numerical simulation of quantum dynamics
(\ref{qmap}) on a classical computer 
is based on the fast Fourier transforms (FFT) between
$\theta$ and $n$ representations. Indeed, the operators
$\hat{U}_k$ and $\hat{U}_T$ are diagonal in $\theta$ and $n$ 
representations respectively, that takes $O(N)$ multiplications for
their realization. The transition between representations
is done by forward and back FFT with $O(N\log_2 N)$ multiplications.
Thus the FFT is the most time consuming part in the classical 
computations of model (\ref{qmap}).

On the contrary the quantum computer requires only
$O(\log^2_2 N)$ gate operations to perform QFT  
(see \cite{shor,ekert96}) and makes 
very easily the forward/back transformations
between $\theta$ and $n$ representations.
Hence, in this part the quantum computer has
the exponential gain comparing to the classical one.
However, it is not so easy to reach the exponential
gain in the multiplication by the diagonal matrices
$\hat{U}_k$ and $\hat{U}_T$ in $\theta$ and
$n$ representations respectively. Of course,
as for the classical computation this can be done
in $O(N)$ operations. In this worst case the quantum computer
will have $ O(\log_2 N)$ gain comparing to the classical
one. We  suppose that a much better 
performance can be reached for the above diagonal part of quantum
algorithm with a strong gain increase. However, in this
paper we leave the question about
maximal gain for future research and 
assume that the unitary diagonal parts 
$\hat{U}_k$ and $\hat{U}_T$ of transformation (\ref{qmap})
are performed by some quantum circuit exactly 
while imperfections are present only in the QFT part.
Namely, for the QFT description in \cite{ekert96} (see Eqs. (14-21) there)
each basic unitary operation $A_j$ (one-qubit) or $B_{jk}$ 
(two-qubit) is rotated on a small random angle of amplitude 
$\epsilon \ll 1$.  At $\epsilon=0$ the
operation $A_j$ is written as $\hat{n_0}\cdot \vec{\sigma}$,
where $\hat{n}_0 = (1/\sqrt{2},0,1/\sqrt{2})$ and $\sigma_i$'s are Pauli 
matrices while with imperfection
$A_j = \hat{n}_j\cdot \vec{\sigma}$  is
achieved by choosing a unit vector $\hat{n}_j$ tilted by an 
angle $\epsilon_j$ ($\leq \epsilon$) from $\hat{n}_0$.  For $B_{jk}$, 
we simply add a random angle of size $\epsilon_{jk}$ 
($|\epsilon_{jk}| \leq \epsilon$) to $\theta_{jk}$ in Eq. (18) of 
\cite{ekert96}. 
These random $\epsilon$-rotations vary in time that produces
an effective noise in the QFT and quantum computing.
In this way the quantum computation with $n_q$ qubits 
models the  kicked rotator dynamics over $N=2^{n_q}$ levels. 
\vskip -0.3cm
\begin{figure}
\epsfxsize=3.2in
\epsfysize=2.6in
\epsffile{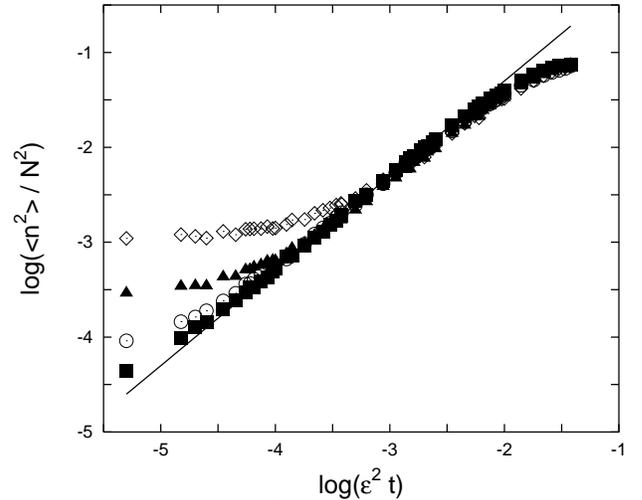}
%\vglue -0.3cm
%\medskip
\caption
{Scaling of $\langle n^2 \rangle / N^2$ is shown for 
various values of $\epsilon$ and $N=2^{n_q}$: $10^{-4} \leq 
\epsilon \leq 2\times 10^{-3}$ and $n_q$ = 10 $(\diamond)$,
11 (full triangle), 12 $(\circ)$, 13 (full square) for $k=10$ 
and $K=5$.  Each point represents the averaged value over $10^3$ kicks 
for $t \leq 10^4$ and
the straight line represents the scaling given by $\langle n^2 \rangle =
D_{\epsilon} t \approx 5 \epsilon^2 N^2 t$ for $t_q < t < t_{\epsilon}$
(see text). Here and in the next figures the logarithms are decimal.
}
\label{fig2}
\end{figure}

The effect of imperfections in the QFT on the second
moment $\langle n^2\rangle$, computed from the probability distribution 
$W_n$ over unperturbed levels $n$ ($\langle n^2\rangle= \sum_n n^2 W_n$),
is shown in Fig.~1 for the regime of quantum chaos ($k > K > 1$)
and different number of qubits $n_q$ at 
$\epsilon=10^{-4}$ and $\epsilon =0$. The data show that the noise from 
imperfections produces an effective diffusive growth of the 
second moment with the rate $D_{\epsilon}$ which grows 
{\it exponentially} with the number of qubits.
In fact the data in Fig.~2 show that in the regime $k > K > 1$
this rate is well described by the relation
$D_{\epsilon} \approx 5 \epsilon^2 2^{2 n_q}$ 
for different $\epsilon, n_q$ and $k$\cite{com}.  
%\vskip -0.3cm
\begin{figure}
\epsfxsize=3.2in
\epsfysize=4.5in
\epsffile{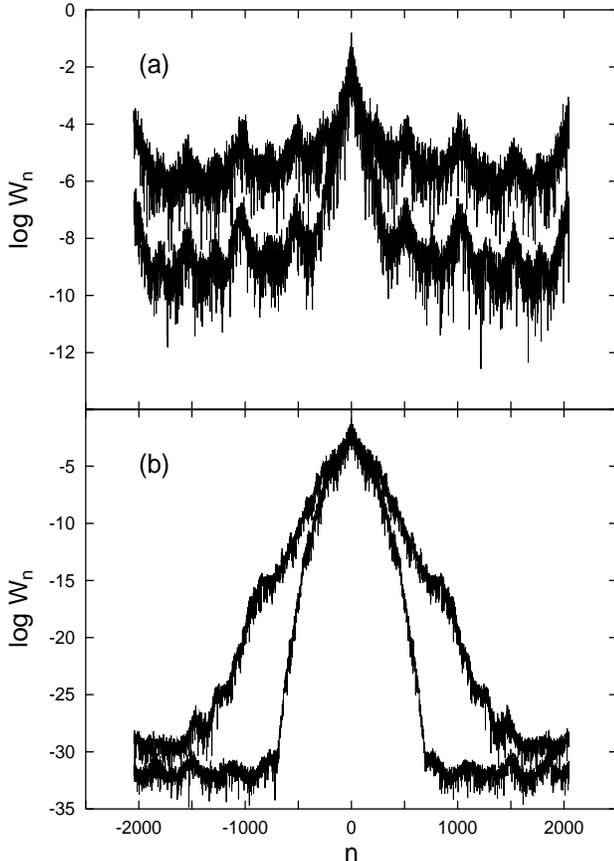}
%\vglue -0.3cm
%\medskip
\caption{Probability distribution $W_n$ over
unperturbed levels for $n_q=12$, $ k=10$ and $ K=5$ 
at two moments of time $t=100$ (lower one) and $t=10^5$ 
(upper one):
(a) $\epsilon = 10^{-4}$ and (b) $\epsilon = 0$. 
Initially all probability is at $n=0$.
}
\label{fig3}
\end{figure}

The physical origin of the exponential error growth in  $\langle n^2\rangle$
with $n_q$ becomes clear from Fig.~3(a)
which shows the probability distribution $W_n$
at two moments of time. At $\epsilon =0$ 
the probability decays exponentially from the 
initially excited level $n=0$ due to dynamical 
localisation (see Fig.~3(b)). This decay continues up to 
plateau with level $W_p  \sim 10^{-32}$ which is determined by the accuracy 
of round-off errors in the classical computer being 
around $\epsilon_c \sim 10^{-16}$. In fact these errors produce 
an effective diffusive growth so that
$W_p \sim {\epsilon_c}^2 t$ and $W_p$
is increased approximately by factor $10^3$ 
when $t$ is changed from 100 to $10^5$.
The classical errors have certain similarities with
the imperfection effects in the quantum case (Fig.~3(a)).
Indeed, the quantum errors also form a plateau $W_{pq}$
at large $n$  the level of which grows diffusively with
time: $W_{pq} \propto \epsilon^2 t$. 
However, the quantum plateau produced by imperfections
has certain differences comparing to the case of a classical
computer. Namely, it is formed
from  clearly pronounced peaks located around
the levels $n_m = \pm 2^m$ with $m=1,2,..,n_q/2$. 
This property is related to the QFT
structure which due to imperfections generates
transitions to $n_q$ levels $n_m$ with probability
$W_{n_m} \propto \epsilon^2$. At $k \ll 1$ the probability
on the levels $n \neq n_m$ is much
smaller than at $n_m$ (data not shown), 
but in the regime of quantum chaos with
$k \gg 1$   each peak starts to take the form
of the central peak with exponential localisation
as at $\epsilon =0$. If the localisation length $l$
is larger than the distance between nearby peaks then
they start to overlap giving more homogeneous 
probability distribution on the plateau. 
In addition these peaks,
as well as the central one at $n=0$,
broaden by imperfection noise. Also between the initial $n_m$ peaks
appear secondary peaks which are placed at powers of 2.
%\vskip -0.5cm
\begin{figure}
\epsfxsize=3.2in
\epsfysize=2.6in
\epsffile{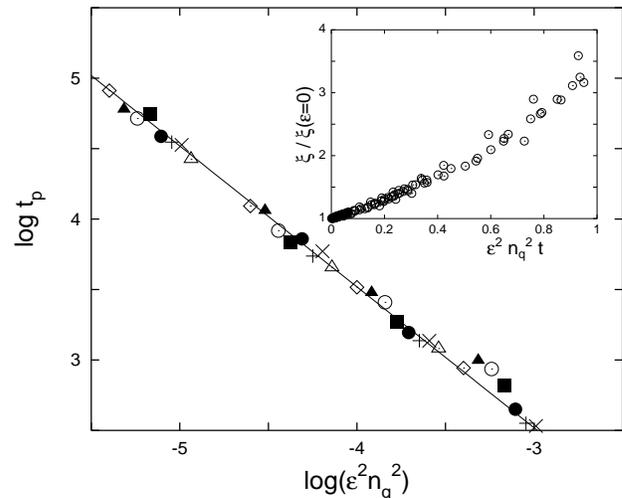}
%\vglue -0.3cm
%\medskip
\caption{Dependence of time scale $t_p$
on system parameters for $ 10^{-4} \leq
\epsilon \leq 2\times 10^{-3}$, 
and $n_q$ = 10 $(\diamond)$,
11 (full triangle), 12 $(\circ)$, 13 (full square), 14 $(\bullet)$,
15 (+), 16 $(\times)$ and 17 $(\triangle)$
for $k=10$ and $K=5$.  The straight line is given by Eq. (\ref{tp}).
Inset shows scaling 
of normalized IPR ratio $\xi(\epsilon)/\xi(\epsilon=0)$.
}
\label{fig4}
\end{figure}

In fact QFT is performed by $O({n_q}^2)$ gate operations
\cite{ekert96} with imperfection rotations. This imperfection noise
creates $n_q$ peaks with probability $W_{pq} \sim \epsilon^2 n_q t$ in each
peak \cite{com1}. 
This leads to the diffusive growth 
$\langle n^2 \rangle \sim N^2 W_{pq} \sim D_{\epsilon} t$
with $D_{\epsilon} \approx n_q \epsilon^2 2^{2n_q}/2$.
The numerical factor here is taken from the data in Fig. 2.
There the variation of $n_q$ by $30\%$ is too small to allow
to distinguish $n_q$ prefactor in front of the exponential dependence
$2^{2n_q}$ from a constant. Since at $\epsilon=0$ the second moment is bounded 
due to quantum localisation of chaos and fluctuates 
around  $\langle n^2\rangle \approx D^2 \approx 4 l^2 \approx k^4/4$ 
\cite{chirikov89,izrailev},
the imperfections strongly modify $\langle n^2\rangle$ after the time scale
\begin{eqnarray} 
\label{tq}
t_q \approx D^2/D_\epsilon \approx k^4/(\epsilon^2 n _q 2^{2 n_q})
\end{eqnarray}
which drops exponentially with $n_q$.
Due to the finite system size, 
the imperfection induced diffusive growth of $\langle n^2\rangle $
is saturated around the maximal value $N^2$ after the 
time $t_{\epsilon} \approx N^2/D_{\epsilon} \approx 2/(n_q \epsilon^2)$
(it is seen in Fig. 2 for large $\epsilon^2 t$).
The imperfection induced diffusion 
exists on the large time interval $t_q \ll t \ll t_{\epsilon}$
(see Fig. 2).

From the above estimates for $W_{pq}$ growth with time it
follows that the probability in $n_q$ peaks becomes comparable with 
the probability inside the central peak  ($W_{pq} n_q 
\sim {n_q}^2 \epsilon^2 t \sim 1$)
after time $t_p \sim 1/({n_q} \epsilon)^2$.
For $t \ll t_p$ the plateau level $W_{pq}$ is rather low
and some characteristics should remain
close to their values in absence of imperfections.
One of them is the inverse participation ratio (IPR) $\xi$
which is often used in the problems with
localisation and determines the effective
number of basis states contributing to the 
wave function. It is defined as $\sum_n {W_n}^2 = 1/\xi$.
By comparing the value of IPR in the presence of
imperfections with its value at $\epsilon =0$
we determine the time scale $t_p$
by the condition $\xi(\epsilon) / \xi(\epsilon=0) =1.5$
for different $n_q, \epsilon$ and $k$.
The dependence of $t_p$ on the parameters 
is shown in Fig. 4 for $k=10$. According to these data 
\begin{eqnarray} 
\label{tp}
t_p \approx 0.33/(\epsilon n_q)^2
\end{eqnarray}
in agreement with the above estimate.
Thus, the dependence of $t_p$ on $n_q$
is polynomial.
We also checked that the numerical
coefficient $C=0.33$ in (\ref{tp}) does not vary
significantly with $k$, e.g. $C \approx 0.32$ for $k=15$
and $C \approx 0.35$ for $k=20$. In our opinion this
is related to the fact that the peaks at $W_{n_m}$
are rather sparse and for large $n_m$ the distance between
them is much larger than 
the localization length $l \approx k^2/4$.
The main consequence of the relation (\ref{tp})
is that certain physical characteristics, e.g. the IPR $\xi$,
remain non-sensitive to imperfections
in the quantum computing during polynomially long
times.

In conclusion, our studies of imperfection effects
on quantum computing of the kicked rotator show
that for certain characteristics,
e.g. the second moment of the probability
distribution, the errors
grow exponentially with the number of qubits $n_q$.
At the same time there are other characteristics, e.g. IPR $\xi$,
which are much more stable and for which the 
errors grow with $n_q$ only polynomially.
However, such stable to imperfections
characteristics are essentially local and are determined only
by a small fraction of levels of the whole Hilbert 
space $N=2^{n_q}$. In a sense the imperfections
determine the precision of quantum computations
and have close similarity with the effect
of round-off errors in the classical computer.
Somehow the quantum computer with imperfections
reminds a {\it very fast} classical computer
with not very high precision. Such a property
can become rather restrictive for certain computations.
At the same time it is possible that the further
development of quantum error-correction codes\cite{steane}
will allow to reach a sufficiently 
high precision of quantum computations.
Also it is known that for the classical computer 
the computational efforts grow only polynomially with the 
mantissa length that can find certain
applications for quantum computing.


\vskip -0.3cm
\begin{thebibliography}{99}
\bibitem{steane} A.~Steane, Rep. Progr. Phys. {\bf 61},
                 117 (1998).
\bibitem{shor} P.~W.~Shor, in Proc. 35th Annu. Symp. Foundations of
                 Computer Science (ed. Goldwasser, S. ), 124 
                 (IEEE Computer Society, Los Alamitos, CA, 1994).
\bibitem{grover} L.~K.~Grover, Phys. Rev. Lett. {\bf 79}, 325 (1997).
\bibitem{zoller} J.~I.~Cirac and P.~Zoller,
                 Phys. Rev. Lett. {\bf 74}, 4091 (1995).
\bibitem{sor2}A.~S\o rensen and K.~M\o lmer,
                 Phys. Rev. Lett. {\bf 82}, 1971 (1999).
\bibitem{nmr} N.~A.~Gershenfeld and I.~L.~Chuang, 
                 Science {\bf 275}, 350 (1997); 
                 D.~G.~Cory, A.~F.~Fahmy and T.~F.~Havel,
                 In Proc. of the 4th Workshop on Physics and Computation
                 (Complex Systems Institute, Boston, MA, 1996).
\bibitem{vagner} V.~Privman, I.~D.~Vagner and G.~Kventsel, 
                 Phys. Lett. A {\bf 239}, 141 (1998).
\bibitem{kane} B.~E.~Kane, Nature {\bf 393}, 133 (1998).
\bibitem{bowden} C.~D.~Bowden and S.~D.~Pethel, Int. J. of Laser Phys., to
                 appear (2000), (quant-ph/9912003).
\bibitem{loss} D.~Loss  and D.~P.~Di~Vincenzo,
                 Phys. Rev. A {\bf 57}, 120 (1998).
\bibitem{cooper} Y.~Nakamura, Yu.~A.~Pashkin, and J.~S.~Tsai,  
                 Nature {\bf 398}, 786 (1999).
\bibitem{lattice} G.~K.~Brennen, C.~M.~Caves, P.~S.~Jessen and I.~H.~Deutsch
                 Phys. Rev. Lett. {\bf 82}, 1060 (1999); 
                 D.~Jaksch, H.~J.~Briegel, 
                 J.~I.~Cirac, C.~W.~Gardiner and P.~Zoller,
                 Phys. Rev. Lett. {\bf 82}, 1975 (1999).
\bibitem{helium} P.~M.~Platzman and M.~I.~Dykman, Science {\bf 284}, 1967
                 (1999).
\bibitem{monroe} C.~Monroe, D.~M.~Meekhof, B.~E.~King, W.~M.~Itano
                 and  D.~J.~Wineland, Phys. Rev. Lett. {\bf 75}, 4714 (1995).
\bibitem{3q} L.~M.~K.~Vandersypen, M.~Steffen, M.~H.~Sherwood,
                 C.~S.~Yannoni, G.~Breyta and I.~L.~Chuang, 
                 Appl. Phys. Lett. {\bf 76}, 646 (2000).
\bibitem{paz} C.~Miquel, J.~P.~Paz and R.~Perazzo, Phys. Rev. A {\bf 54},
                 2605 (1996).
\bibitem{zurek} C.~Miquel, J.~P.~Paz and W.~H.~Zurek, 
                 Phys. Rev. Lett. {\bf 78}, 3971 (1997).
\bibitem{gs} B.~Georgeot and D.~L.~Shepelyansky, quant-ph/9909074;
                 quant-ph/0005015; D.~L.~Shepelyansky, quant-ph\-/0006073.
\bibitem{fishman} S.~Fishman, D.R.~Grempel, R.E.~Prange,
                 Phys. Rev. Lett. {\bf 49}, 509  (1982).
\bibitem{chirikov89} B.~V.~Chirikov, in {\it Chaos and Quantum Physics}, 
                 Les Houches Lecture Series  52,
                 Eds. M.-J. Giannoni, A.Voros, and J. Zinn-Justin 
                 (North-Holland, Amsterdam, 1991), p. 443.
\bibitem{izrailev} F.M.~Izrailev, Phys. Rep. {\bf 129}, 299 (1990).
\bibitem{ds83} D.~L.~Shepelyansky, Physica D {\bf 8}, 208 (1983).
\bibitem{ott}  E.~Ott, T.~M.~Antonsen, and J.~D.~Hanson 
                  Phys. Rev. Lett. {\bf 53}, 2187 (1984).
\bibitem{casati79} G.~Casati, B.~V.~Chirikov, J.~Ford and F.~M.~Izrailev,
                  Lecture Notes Phys. {\bf 93}, 334 (1979).
\bibitem{raizen} F.~L.~Moore, J.~C.~Robinson,
               C.~F.~Bharucha, B.~Sundaram and M.~G.~Raizen, 
               Phys. Rev. Lett. {\bf 75}, 4598 (1995).                         
\bibitem{amman} H.~Ammann, R.~Gray, I.~Shvarchuck and
               N.~Christensen, Phys. Rev. Lett. {\bf 80}, 4111 (1998).  
\bibitem{ekert96} A.~Ekert And R.~Jozsa, Rev. Mod. Phys. {\bf 68}, 733 (1996).
\bibitem{com} We checked that the data for $k=15, \; 20$ and $K=5$ are also 
              consistent with the expression 
              $D_{\epsilon}  \approx 5 \epsilon^2 N^2 $.
\bibitem{com1} Noise amplitude is effectively increased
              by a factor $\sqrt{n_q}$ since $O(n_q)$ gate operations 
              are performed per each peak.

\end{thebibliography}
\end{document}